\documentclass[preprint]{aastex}

\newcommand{\xsv}{\sigma^2_{xs}}
\newcommand{\et}{et al.\ }

\newcommand{\ls}
{\mathrel{\hbox{\rlap{\hbox{\lower4pt\hbox{$\sim$}}}\hbox{$<$}}}}
\newcommand{\gs}
{\mathrel{\hbox{\rlap{\hbox{\lower4pt\hbox{$\sim$}}}\hbox{$>$}}}}

\newcommand{\xmm}{{\it XMM-Newton}}

\newcommand{\sax}{{\it BeppoSAX}}

\newcommand{\asca}{{\it ASCA}}
\newcommand{\pks}{PKS~2155--304}

\begin{document}

\title{ High Temporal Resolution \xmm\ Monitoring of \pks}

\shorttitle{ \xmm\ Monitoring of \pks}
\shortauthors{ Edelson \et }

\author{Rick Edelson\altaffilmark{1,2},
        Gareth Griffiths\altaffilmark{2},
        Alex Markowitz\altaffilmark{1},
        Steve Sembay\altaffilmark{2},
        Martin J.\ L.\ Turner\altaffilmark{2},
        Robert Warwick\altaffilmark{2} }

\email{rae@astro.ucla.edu}

\altaffiltext{1}{Astronomy Department; University of California; Los
Angeles, CA 90095-1562; USA}

\altaffiltext{2}{X-ray Astronomy Group; Leicester University; Leicester
LE1 7RH; United Kingdom}

\begin{abstract}

The bright, strongly variable BL Lac object \pks\ was observed by \xmm\
for two essentially uninterrupted periods of $\sim$11 and 16~hr on 30-31
May 2000.
The strongest variations occurred in the highest energy bands.
After scaling for this effect, the three softest bands
(0.1--1.7~keV) showed strong correlation with no measurable lag to
reliable limits of $ | \tau | \ls 0.3 $~hr.
However, the hardest band ($\sim$3~keV) was less well-correlated with the
other three, especially on short time scales, showing deviations of 
$\sim$10--20\% in $\sim$1~hr although, again, no significant interband lag
was detected.
This result and examination of previous \asca\ and \sax\ cross-correlation
functions suggest that previous claims of soft lags on time scales of 
0.3--4~hr could well be an artifact of periodic interruptions due to 
Earth-occultation every 1.6~hr.
Previous determinations of the magnetic field/bulk Lorentz factor were
therefore premature, as these data provide only a lower limit of 
$ B \gamma^{1/3} \gs 2.5 $~G.
The hardest band encompasses the spectral region above the high-energy
break; its enhanced variability could be indicating that the break energy
of the synchrotron spectrum, and therefore of the underlying electron
energy distribution, changes independently of the lower energies.

\end{abstract}


\section{ Introduction }

Blazars (BL Lacertae objects and optically violently variable
quasars) show strong and rapid $\gamma$-/X-ray variability.
When combined with their observed large apparent luminosities, this
indicates that the emission we observe is beamed towards us, as isotropic
emission would have to be highly super-Eddington (e.g., Bregman 1990).
Likewise, determining the interrelations between variations in different
energy bands can define geometrical and causal relations between emission
regions and processes.
For instance, X-ray variations in the blazar \pks\ have been seen to lead
those in the optical/ultraviolet (e.g., Edelson \et 1995, Urry \et 1997),
indicating that the X-rays are not produced by Compton scattering from
these lower-energy bands.

In $\sim$2~day long \sax\ and \asca\ monitoring of \pks, Chiappetti \et
(1999), Zheng \et (2000) and Kataoka \et (2000) reported that variations
in the soft X-ray band (0.1--1.5~keV) lagged behind those in the hard
X-ray band (3.5--12~keV) by $\sim$0.3--4~hr.
These ``soft lags" were explained as due to energy-dependent radiative
cooling losses.
Such small interband lags have been claimed in monitoring of Mkn~421, a
similar type of blazar: Takahashi \et (1996) found soft lags, Fossati \et
(2000) found hard lags, and Takahashi \et (2000) found both hard and soft
lags in a single very long observation.
However, all of those observations were made with low-Earth orbit
satellites, for which the input light curves were corrupted by
Earth-occultation interruptions every $\sim$1.6~hr orbit, making highly
problematic the unambiguous detection of short lags as discussed in
\S~4.1.

This paper reports the results of essentially continuous $\sim$11 and
16~hr \xmm\ observations of \pks.
With its highly eccentric $\sim$48~hr orbit (permitting long uninterrupted
viewing), high throughput and wide bandpass, \xmm\ is ideally suited for
reliably measuring small interband lags.
The observations and data reduction are discussed in the next section, the
temporal analysis is reported in \S~3 and the results are discussed in
\S~4. 

\section{ Observations and Data Reduction }

\subsection{ Observations }

\pks\ was observed during \xmm\ orbit \# 87, from 2000 May 30 05:28:29 UT
to 2000 May 31 20:40:34 UT.
Because this performance verification/guaranteed time observation occurred
early in the mission, only a fraction of the currently available orbit was
spent gathering data, and some of the instruments were in non-standard
configurations that precluded time-series analysis.
\xmm\ (Jansen \et 2001) made useful observations of \pks\ during two
uninterrupted intervals, ``Interval A," which occurred before the
$\sim$4~hr mid-orbit interruption, and ``Interval B," which occurred
afterwards.
Observation logs are given in Table~1.

\placetable{table1}

 Because this experiment required very high sensitivity in order to reach
acceptable signal-to-noise ratio in the broadband flux in the shortest
time and thus be sensitive to the smallest possible interband lags, only
data from the high-throughput European Photon Imaging Camera
(EPIC) instruments were considered.
One of the EPIC instruments contains pn CCDs (Struder \et 2001) while the
other two comprise MOS CCDs (Turner \et 2001).
During Interval A, MOS1 was operated in partial window mode (PRI PART
W4) and MOS2 in timing mode.
During Interval B, MOS1 was operated in timing mode and MOS2 in PRI PART
W4.
In PRI PART W4 mode, only the central 100 x 100 pixels of the MOS central
chip is read out, giving a frame integration time of $\sim$0.15~s and
minimizing the effects of pileup.
Also, the pn was in Small Window mode for both intervals, yielding an
integration time of $\sim$ 5.7 ms, again in order to prevent pileup.
The medium filter was selected for all three instruments throughout.
Due to current uncertainties in the calibration this paper does not
consider timing mode data, so the Interval A MOS2 and Interval B MOS1 data
were ignored.

\subsection{ Data Reduction }

A standard reduction of raw pn and MOS event lists was performed using the
Science Analysis System (SAS).
This involved the subtraction of hot, dead or flickering pixels, removal
of events due to electronic noise and correction of event energies for
charge transfer losses.
For the pn, only pattern 0 events were used while events with patterns
0-12 were selected for the MOS cameras.
The pattern of an event defines the shape of the charge cloud created by
that event, similar to \asca\ ``grades," so the smaller pattern number
means
a more concentrated and therefore more reliable event (e.g., pattern 0
corresponds to a single pixel event).

X-ray light curves were accumulated for all three cameras from events
extracted within a circle of radius 45$^{\prime\prime}$ around the source
centroid.
No background light curve was measured since the broadband source count
rates are of order $\sim$1000 times the background.
Exposure times of approximately 11~hr were obtained during Interval A and
16~hr during Interval B.
The data were initially extracted in 3~s intervals, then rebinned to
0.167~hr (600 s) bins.
At about May 30 19:15 and again at May 31 10:00, the pn shut down for
$\sim$0.1~hr; values for these two points were derived by linear
interpolation from the adjacent points.
Light curves were extracted for the entire usable band (0.1-12 keV).
The data were also broken up into four bands of 0.1-0.35 keV, 0.4-0.75
keV, 0.9-1.7 keV and 2.0-12.0 keV.
These light curves, called VS (e.g., Very Soft), S, H and VH,
respectively, form the basis of the analysis below.

\section{ Comparison of Hard and Soft Band Light Curves }

The light curves are shown in Figure~1.
The variability is lower than seen in some historical campaigns, but some
of these were triggered e.g., by $\gamma$-ray flares, so they might be
expected to show enhanced variability.
Because \pks\ is so bright in the EPIC-pn ($\sim$60~ct/s), significant
variations are seen above the $\sim$1\% errors down to very short time
scales ($\sim$0.3~hr).

\placefigure{fig1.ps}

\subsection{ Variability Amplitudes }

An obvious feature of Figure~1 is that the strongest variability is seen
in the hardest bands.
The strong relation between energy band and variability level is confirmed
in Table~2.
The excess variance is defined as 
$ \xsv =  ( S^2 - \langle \sigma_{err}^2 \rangle ) / \langle X \rangle^2
$, where $S^2$ is the measured variance of the light curve and $ \langle X
\rangle $ and $ \langle \sigma_{err}^2 \rangle $ are the mean and mean
square errors, respectively).
The more intuitive fractional variability parameter is simply the square
root of the excess variance: $ F_{var} = \sqrt{\xsv} $.
Note that $F_{var}$ increases monotonically and strongly with energy band,
e.g., from 1.5\% at $\sim$0.2~keV to 5.4\% at $\sim$3~keV in Interval~B.

The Interval~B data are more useful because they cover a $\sim$50\% longer
time span and show stronger and sharper variations than Interval~A.
Also, the pn count rate is $\sim$2.6 times the MOS count rate and direct
comparison of scaled pn and MOS data from the same energy bands and
intervals showed excellent agreement within the errors.
Thus, only the pn data are used in the following analysis, and the MOS are
not used.

\placetable{table2}

\subsection{ Cross-Correlation Analysis }

The first step in the construction of a cross-correlation function
(CCF) is to normalize the two light curves to zero mean and unit variance
by subtracting the mean and dividing by the rms of the light curve.
Plots scaled in this fashion provide an intuitive way to visualize the
agreement between two light curves, as shown for two pairs of light curves
in Figure~2.

\subsubsection{ Correlations and Divergences Between the Light Curves }

Figure~2 presents the most and least well-correlated pairs of light curves
in this study.
The H and S band light curves are essentially identical within the errors
(Figure~2, top), and in fact all of the softest bands (0.1--1.7~keV) track
well.
However, the $\sim$3~keV VH band shows significant divergences, as is most
clearly seen in the plot with the VS band (Figure~2, bottom).
The VH band shows flare-like features that are not seen in other bands,
with durations of $\sim$1~hr, at $\sim$10:00--12:00 of Interval~A and
$\sim$08:30--10:30 in Interval~B.
However, on longer time scales, the VH band does appear to track the three
softer bands.
Throughout all the light curves, there is no obvious tendency for
variations in one band to lead those in another.

\subsubsection{ Interband Lags }

CCFs were measured using both the interpolated (ICF; White \& Peterson
1994) and the discrete correlation functions (DCF; Edelson \& Krolik
1989).
Uncertainties in the ICF lags were estimated using the bootstrap method of
Peterson \et (1998).
The CCF results are tabulated for all pairs of bands in Table~3, with the
data ordered by descending $r_{max}$ in the (better-defined) Interval~B
CCFs.

\placetable{table3}

The CCFs shown in Figure~3, derived from the data in Figure~2, were chosen
to cover the full range of data quality in this experiment.
In the most useful interval (Interval~B), the best-correlated pair (H vs.\
S) show a classic sharp, single-peaked CCF, with a large correlation
coefficient ($ r_{max} \approx 0.9 $).
These data are ideal for measuring lags, and in fact the measured lag is
zero to within the errors.
The formal 90\% confidence limit is $ | \tau | \ls 0.12 $~hr, but we 
assign a more conservative minimum 2$\sigma$ limit of twice the mean
sampling bin, or in this case, 0.3~hr.
We note that the other two soft CCFs (VS vs.\ S and VS vs.\ H) are also
consistent with $ | \tau | \ls 0.3 $~hr, so we assign this limit to
Interval~B lags throughout the soft bands.
In Interval~A, the light curve is shorter and the variations are not as
sharp, and the peak is correspondingly less well-defined, with limits of
only $ | \tau | \ls 0.6 $~hr throughout the soft bands.
In all but one of these cases, the mean lags are within the 90\%
confidence limits and of the 12 runs (six with the DCF and six with the
ICF), five were positive, five were negative, and two were identically
zero.
Thus, we conclude that there is no evidence for soft or hard lags within
the 0.1--1.7~keV band, to conservative limits of $ | \tau | \ls 0.3 $~hr
in Interval~A and $ | \tau | \ls 0.6 $~hr in Interval~A.

\placefigure{fig3.ps}

Somewhat different behavior may be indicated by the VH vs.\ VS CCF.
The Interval~B data are not very strongly correlated ($ r_{max} \approx
0.64 $), but they do show a well-defined albeit broad CCF peak.
The limit on the interband lag is correspondingly worse, $ | \tau | \ls 1
$~hr, and similarly weak limits are derived for the VH/S and VH/H CCFs.
The Interval~A CCF is even worse, showing two essentially equal broad
peaks at about $\pm$1~hr, with a correlation coefficient of $ r_{max}
\approx 0.57 $).
The formal limit on this CCF is also $ | \tau | \ls 1 $~hr.
Again, the ICF lags are within than the 90\% confidence limits and of the
12 ICF and DCF runs, four were positive, seven were negative, and one
identically zero.
More importantly, the Inverval~A VH/VS CCF clearly fails a common-sense
visual inspection of the CCF as neither peak is statistically more
significant than the other. 
It is essentially unusable for time series analysis.
That is, in the absence of the Interval~B data, no conclusions whatsoever
could be reached with the Interval~A data alone. 

\section{ Discussion }

In this set of two uninterrupted \xmm\ long-looks at \pks, no evidence was
found for soft interband lags (or for hard lags).
The variability amplitude increases dramatically from soft to hard
energies.
After scaling for these differences, the soft band (0.1--1.7~keV) light
curves track well, but the very hard band ($\sim$3~keV) light curve shows
small but significant divergences with the soft bands on shorter time
scales.

\subsection{ Comparison with Previous Observations }

The lack of an interband lag appears to contradict previous claims of
$\sim$0.3--4~hr lags, based on monitoring by the low-Earth orbit
telescopes \asca\ in 1994 and \sax\ in 1996 and 1997.
However, those $\sim$2~day observations had only $\sim$50\% (or
worse) duty cycles, suffering from Earth-occultation interruptions at the
1.6~hr orbital period that tremendously complicate the measurement of
short lags.
The uninterrupted, high signal-to-noise \xmm\ data do not have these
problems.

These problems are clearly seen in the 1997 November \sax\ observation.
Chiappetti \et (1999) found that variations in the hard band
(3.5--10~keV) led those in the soft band (0.1--1.5~keV) by $ 0.5 \pm 0.08
$~hr. 
However, the CCF (see Figure~3 of Chiappetti \et 1999) shows a strong,
periodic feature, with very large errors, large negative deviations,
and/or missing data at, e.g., --5.2, --3.6, --2.1, --0.5, +1.1, +2.7,
+4.3~hr.
These occur at the satellite orbital period ($\sim$1.6~hr), as would be
expected because both light curves suffer interruptions at this
period due to Earth occultation.
(The excursions/interruptions are asymmetric with respect to zero lag,
possibly due to different acceptance criteria for the low and medium 
energy light curves that covered slightly different periods.)
This systematic corruption of the CCF on short time scales makes it highly
problematic to reliably measure lags $\ls$1 orbit.
In this particular case, the asymmetry of the periodic artifact means that
near zero lag, negative lags are more strongly biased down, shifting the
center of weight of the CCF higher and producing the impression of a
positive lag.
Note also that the three highest points in this CCF (excluding the one
very noisy point) are also the three closest to zero (with $ |\tau | \le
0.2 $~hr), and that the claimed lag actually lies near a local minimum in
the CCF.
The same effect is seen the Zheng \et (2000) CCF derived from the same
data (Figure~5a of that paper): a strong, asymmetric artifact with a
1.6~hr period.
Again, although that paper claimed a soft lag of $\sim$0.3~hr, the three
highest points indicate no measurable lag.

Likewise, both Zheng \et (2000) and Kataoka \et (2000) claimed soft lags
of $\sim$0.3--1.1~hr, increasing as the gap between the hard and soft
bands was increased, in the 1994 \asca\ data.
Kataoka \et (2000) did not show a CCF but that in Zheng \et
(2000; Figure~5e) is weak ($ r_{max} \approx 0.6$) and broad and
consistent with any lag between --2 and +4~hr.
Finally, a $\sim$4~hr soft lag was claimed in the 1996 \sax\ CCF (Zheng
\et 2000; Figure~5c), but that correlation is also weak ($ r_{max} \approx
0.7 $) and broad, approximately constant for all lags between 0 and +4~hr.
That light curve was also corrupted by interruptions on longer 
($\sim$4--6~hr) time scales, possibly introducing artifacts on these 
time scales as well.

Interband lags have also been claimed in a similar HBL (see below),
Mkn~421.
Takahashi \et (1996) found soft lags smaller than 1~orbit in an \asca\
campaign, with the same sort of band-gap dependence claimed in the 1994
\asca\ observations of \pks.
Fossati \et (2000) found the opposite: that the hard band lagged behind 
the soft in a series of \sax\ observations.
Finally, in a very long (11-d) \asca\ observation, Takahashi \et
(2000) tentatively claimed that Mkn~421 exhibited both hard and soft X-ray
lags, again on time scales of $\ls$1~orbit, but noted the possibility that
the result could be spurious, caused by periodic Earth-occultation.
CCFs were not shown in Takahashi \et (1996, 2000), but they were in
Fossati \et (2000), and showed the same kind of periodic corruption as the
1997 \pks\ CCFs.
Takahashi \et (2000) found the detection of both hard and soft lags
challenging to explain theoretically, but there is a simple observational
explanation: that none of the lags, hard and soft, shorter than or of
order the orbital interruption time scale, are real. 

In summary, the limits on the interband lag in this campaign appear
inconsistent or only marginally consistent with previous claims of soft
(and hard) lags in this source and Mkn~421.
However, direct examination of those CCFs indicate that they are in fact
fully consistent with zero lag, and that claims of measurable lags could
easily have resulted from periodic artifacts caused by Earth-occultation.

\subsection{ Theoretical Implications }

Blazar SEDs have two distinct components, with one dominating at lower
energies and peaking in the X-ray regime, and the other peaking in
$\gamma$-rays.
This suggests that there are two emission processes operating: incoherent
synchrotron emission at lower energies and Compton scattering at higher
energies.
The possibility that the synchrotron photons could be the seeds for the
Compton emission (the synchrotron self-Compton, or SSC process) is
consistent with the data, but an unambiguous link has not been
established.
SED snapshots have been obtained for many blazars, which appear to
segregate into two classes that may or may not be extrema of a single
population: HBLs, for which the low-energy component peaks in the X-rays
and dominates emission in that band, and LBLs, which peak at lower
(ultraviolet/optical) energies and for which either the synchrotron or
Compton component can dominate in the X-rays.
HBLs tend to show no emission lines (while LBLs do) and to have lower
luminosities than LBLs (e.g., Sikora \et 1997).
Both \pks\ and Mkn~421 (the other blazar for which soft lags are
claimed) are HBLs, and are two of only five AGN (all HBLs) that have been
detected at TeV energies.

A variety of models can fit these single-epoch SEDs (e.g., Ghisellini \et
1998).
These provide a framework for understanding blazar physics, but cannot
fully exploit the potential constraints offered by the strong {\it
variability} seen in blazars.
Time-dependent models are being developed (e.g., Mastichiadas \& Kirk
1998), but as of this writing they have focused on a few special cases and
are not directly applicable to these data.

In the absence of such detailed models, these results still allow
determination of some general constraints on the physical conditions in
\pks.
Measurement of a soft lag could have given the cooling time scale and
therefore an estimate of the magnetic field (assuming a value for the
Lorentz factor $\delta$), assuming that the lag is due to the electron
population softening as the hard electrons preferentially lose energy due
to radiative losses by the following formula (from Chiappetti \et(1999)):
\begin{equation}
B \delta^{1/3} = 300 \bigg({ 1+z \over \nu_1 }\bigg)^{1/3} 
\bigg({ 1 - (\nu_1/\nu_0)^{1/2}\over \tau} \bigg)^{2/3} {\rm G},
\end{equation}
where $B$ is the magnetic field strength,  $z$ is the redshift, and 
$\nu_1$ and $\nu_0$ are the frequencies (in units of $10^{17}$~Hz) at 
which the lag, $\tau$, is measured.
Because this experiment only yielded an upper limit on $\tau$, only a
lower limit can be obtained for $ B \delta^{1/3} $.
For the parameters measured in the H/S CCF, $z = 0.12 $, $\nu_0 = 3 $,
$ \nu_1 = 0.5 $ (1.2 and 0.2~keV, respectively), and $\tau \ls 1000 $~s,
we derive a lower limit of $ B \delta^{1/3} \gs 2.5 $~G.
A similar constraint is obtained with the 1/4 CCF data: the band gap is
larger, with $\nu_0 = 7 $ and $ \nu_1 = 0.5 $ (3 and 0.2~keV,
respectively), but the limit on the lag is much worse, $\tau \ls 4000 $~s,
so a less stringent limit of $ B \delta^{1/3} \gs 1.3 $~G is derived.
These limits are marginally inconsistent with those based on the claimed
detection of an interband lag, e.g., $ B \delta^{1/3} = 2.5 $~G (Kataoka
\et 2000).

Perhaps the most interesting result of this experiment is that the VH
($\sim$3~keV) band shows significant differences with the three softer
bands (0.1--1.7~keV) on time scales of $\ls$1~hr.
The time-averaged spectrum of \pks\ shows a break between these bands,
steepening from $ \Gamma \approx 2.45 $ to $ \Gamma \approx 2.7 $ at about
$\sim$2~keV.
This means that the VH band would be sensitive to changes in the break
energy.
The fact that the variations in this band are much more pronounced than,
and relatively less correlated with those in the softer (power-law) part
of the spectrum could indicate that the peak of the synchrotron spectrum
is changing in a manner unrelated to the changes in the normalization.
As the peak of the spectrum corresponds to the maximum energy in the
power-law distribution of electrons, this could be indicating that the
heating time scale is becoming comparable to the light-crossing time scale
at these energies.
However, this effect needs to be firmly established and explored in
greater detail, e.g., by extended \xmm\ monitoring of this source and
Mkn~421, the other blazar for which short lags have been claimed, and the
brightest in the sky.

\acknowledgments
The authors thank the \xmm\ team, especially at Leicester and Vilspa, for
the smooth operation of this new, complex and very powerful instrument.
They also appreciate Greg Madejski and Laura Maraschi for discussions that
aided greatly their understanding of blazar physics and preparing of the
observing proposal.
Edelson thanks the people of Vietnam for their hospitality while this
paper was being written.
Edelson and Markowitz were supported by NASA grants NAG~5-7317 and
NAG~5-9023.

\begin{deluxetable}{lcc}
\tablewidth{3.5in}
\tablenum{1}
\tablecaption{ Observing Log \label{tab1}}
\small
\tablehead{
\colhead{ Interval } & \colhead{ Instrument } 
& \colhead{ Obs.\ Period (2000 May UT) } }
\startdata
A &  pn   & 30 10:20:09 -- 30 20:53:29 \\
A &  MOS1 & 30 10:58:41 -- 30 21:53:42 \\
B &  pn   & 31 00:52:59 -- 31 17:21:38 \\
B &  MOS2 & 31 01:32:15 -- 31 17:20:29 \\
\enddata
\end{deluxetable}

\begin{deluxetable}{ccccccccc}
\tablewidth{6.5in}
\tablenum{2}
\tablecaption{ Variability Parameters \label{tab2}}
\small
\tablehead{
\colhead{ } & \colhead{ Band } & \colhead{ Band Ctr.} & 
\colhead{ Int.\ A } & \colhead{ Int.\ A } & \colhead{ Int.\ A } & 
\colhead{ Int.\ B } & \colhead{ Int.\ B } & \colhead{ Int.\ B } \\ 
\colhead{ Camera } & \colhead{ (keV) } & \colhead{ (keV)  } & 
\colhead{ ct/s } & \colhead{ $F_{var}$ (\%) } &
\colhead{ $\xsv$ ($\times 10^{-4}$) } & 
\colhead{ ct/s } & \colhead{ $F_{var}$ (\%) } &
\colhead{ $\xsv$ ($\times 10^{-4}$) } }
\startdata
pn  & 0.1--12   & 0.7 & 61.4 & 1.9 &  3.7 & 60.6 & 2.4 &  5.7 \\     
pn  & 0.1--0.35 & 0.2 & 20.5 & 1.1 &  1.2 & 19.8 & 1.5 &  2.4 \\     
pn  & 0.4--0.75 & 0.6 & 18.7 & 1.9 &  3.8 & 18.4 & 2.3 &  5.1 \\     
pn  & 0.9--1.7  & 1.2 & 10.4 & 2.8 &  8.1 & 10.5 & 3.4 & 11.8 \\     
pn  & 2.0--12   &   3 &  2.7 & 4.2 & 17.9 &  2.9 & 5.4 & 28.7 \\     
MOS & 0.1--12   & 0.7 & 23.2 & 2.6 &  6.7 & 23.8 & 2.8 &  7.6 \\     
MOS & 0.1--0.35 & 0.2 &  4.6 & 1.4 &  1.9 &  4.7 & 1.5 &  2.2 \\     
MOS & 0.4--0.75 & 0.6 &  6.1 & 2.5 &  6.1 &  6.2 & 2.1 &  4.5 \\     
MOS & 0.9--1.7  & 1.2 &  6.4 & 3.2 & 10.3 &  6.6 & 3.5 & 12.0 \\     
MOS & 2.0--12   &   3 &  2.1 & 4.4 & 19.3 &  2.3 & 5.1 & 25.5 \\     
\enddata
\end{deluxetable}

\begin{deluxetable}{ccccccc}
\tablewidth{5in}
\tablenum{3}
\tablecaption{ Cross-Correlation Results \label{tab3}}
\small
\tablehead{
\colhead{ Band 1 } & \colhead{ Band 2 } & \colhead{ } &
\colhead{ DCF } & \colhead{ DCF } & \colhead{ ICF } & \colhead{ ICF } \\
\colhead{ (keV) } & \colhead{ (keV) } & \colhead{ Interval } &
\colhead{ $r_{max}$ } & \colhead{ $\tau$ (hr) } & 
\colhead{ $r_{max}$ } & \colhead{ $\tau$ (hr) } }
\startdata
0.4--0.75 & 0.9--1.7  & A & 0.85 &   0.00 & 0.88 &  +0.08 $\pm$ 0.50 \\
0.1--0.35 & 0.4--0.75 & A & 0.74 & --0.83 & 0.77 & --0.75 $\pm$ 0.46 \\
0.1--0.35 & 0.9--1.7  & A & 0.71 & --0.50 & 0.75 & --0.42 $\pm$ 0.69 \\
0.9--1.7  & 2.0--12   & A & 0.78 &  +0.17 & 0.82 &  +0.08 $\pm$ 0.69 \\
0.4--0.75 & 2.0--12   & A & 0.70 &   0.00 & 0.75 &  +0.08 $\pm$ 1.08 \\
0.1--0.35 & 2.0--12   & A & 0.56 &  +0.50 & 0.59 & --0.92 $\pm$ 1.00 \\
0.4--0.75 & 0.9--1.7  & B & 0.89 & --0.17 & 0.91 & --0.08 $\pm$ 0.12 \\
0.1--0.35 & 0.4--0.75 & B & 0.80 &  +0.33 & 0.84 &  +0.25 $\pm$ 0.29 \\
0.1--0.35 & 0.9--1.7  & B & 0.78 &  +0.17 & 0.82 &  +0.25 $\pm$ 0.31 \\
0.9--1.7  & 2.0--12   & B & 0.76 &   0.00 & 0.79 & --0.08 $\pm$ 0.17 \\
0.4--0.75 & 2.0--12   & B & 0.74 & --0.33 & 0.77 & --0.25 $\pm$ 0.64 \\
0.1--0.35 & 2.0--12   & B & 0.62 & --0.50 & 0.65 & --0.25 $\pm$ 1.08 \\
\enddata
\end{deluxetable}

\clearpage

\begin{figure}[h]
\includegraphics{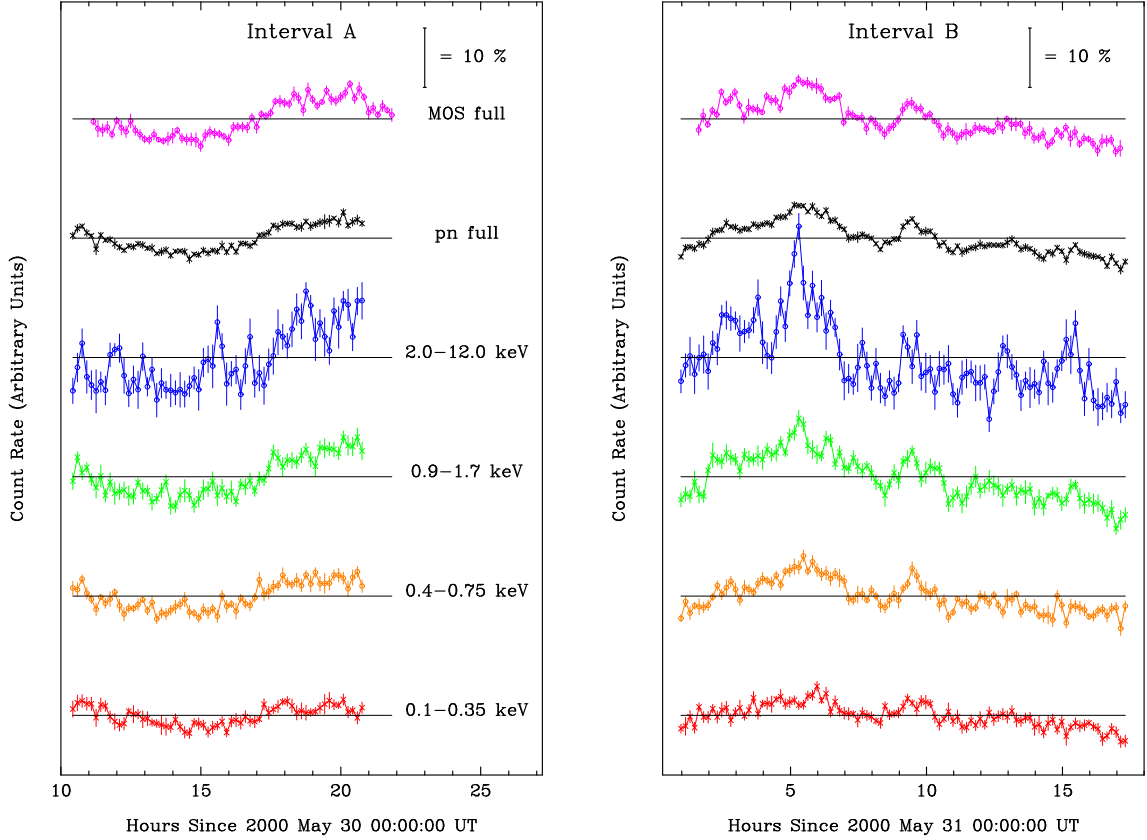}
\vspace{5.5in}
\caption{ 
\xmm\ light curves for \pks, for Interval A (left) and Interval B (right).
The data are binned in 0.167~hr intervals, and a logarithmic y-axis is
used.
From top to bottom, count rates are given for the MOS total (0.1--12~keV),
pn total (0.1--12~keV), and four pn subbands (2.0--12, 0.9--1.7, 0.4--0.75
and 
0.1--0.35~keV, or VH, H, S and VS, respectively). }
\label{fig1}
\end{figure}

\clearpage

\begin{figure}[h]
\includegraphics{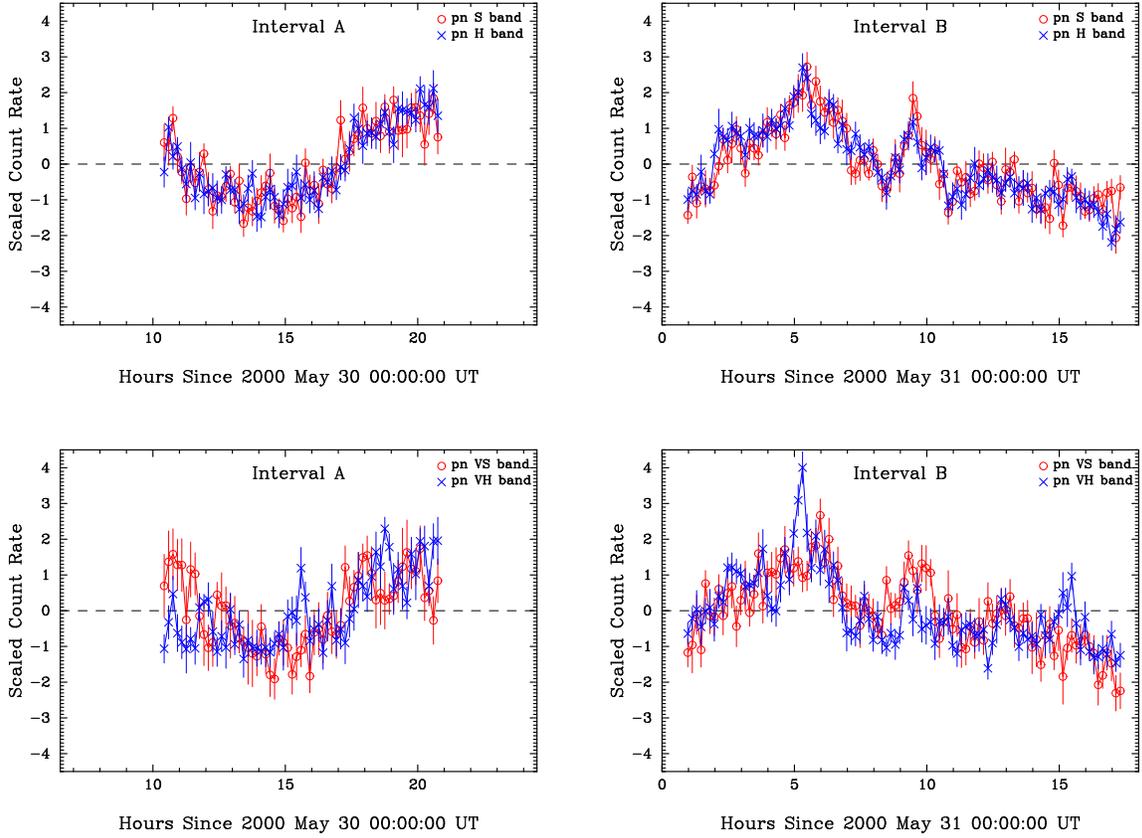}
\vspace{5.5in}
\caption{Scaled \pks\ pn soft (circles) and hard band ($\times$s) light
curves.
Data are shown for bands H/S (top) and VH/VS (bottom), over Interval~A
(left) and Interval~B (right).
The data in each band first had the mean count rate subtracted, and then
were divided by the rms (that is, the standard initial scaling before
performing the CCF), so that they have the same apparent means and
variances.
Note how well the H and S light curves track each other, and that the VH
and VS light curves track well on long time scales, although significant
detailed differences can be seen between the VH and VS bands on short time
scales.}
\label{fig2}
\end{figure}

\clearpage

\begin{figure}[h]
\includegraphics{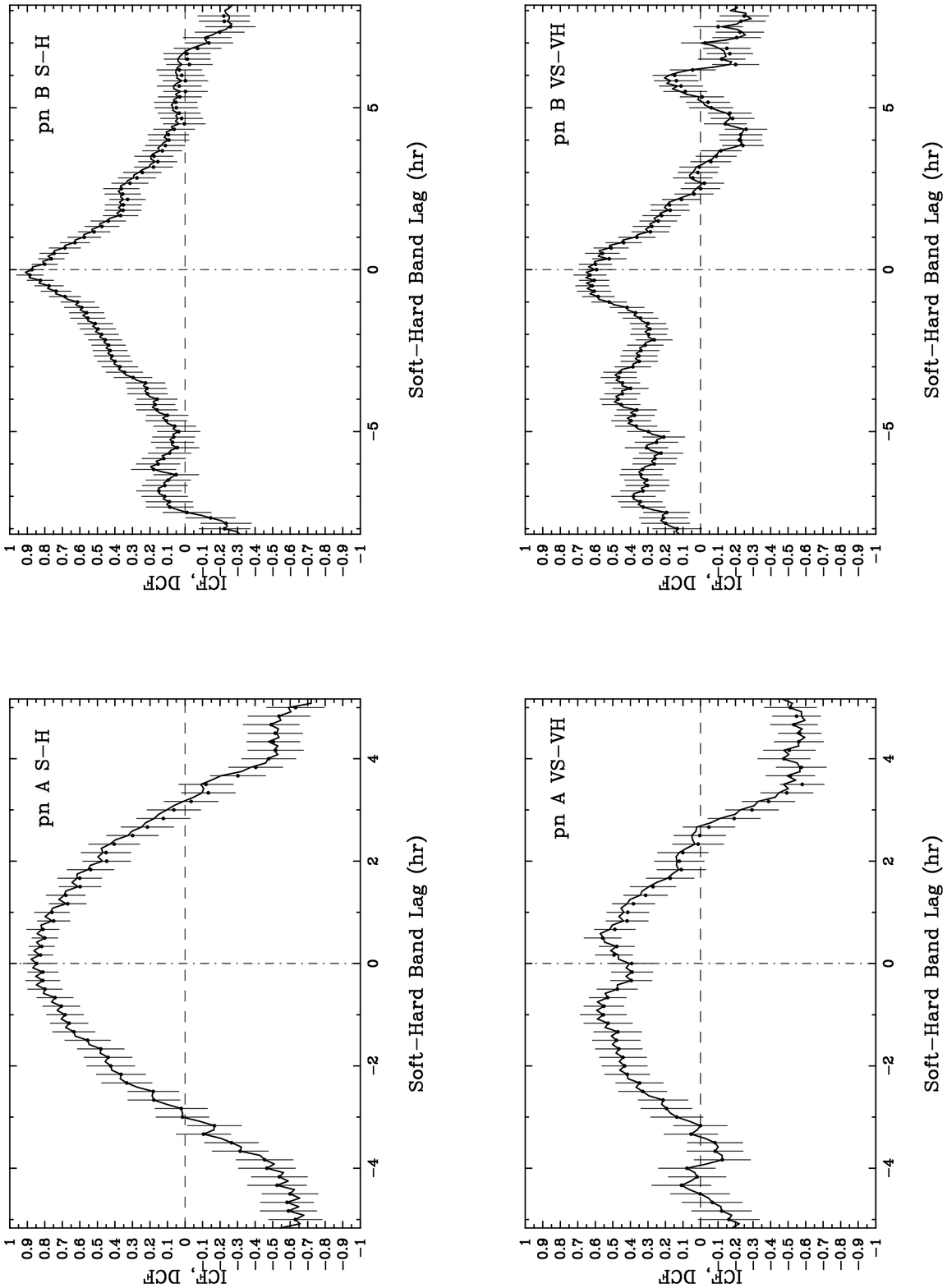}
\vspace{5.5in}
\caption{CCFs for bands H/S (top) and VH/VS (bottom), over Interval~A
(left) and Interval~B (right).
The DCF, shown as circles with error bars, was binned at the natural time
scale of the input light curve, 0.167~hr.
The ICF, shown as a solid line, is binned on a time scale of 0.067~hr.
All of the measured CCFs are consistent with zero lag, with no tendency
for one band to lead the other.
Note the range of quality in the CCFs, with the H/S Interval~B data
showing a strong sharp peak ($r_{max} \approx 0.9$) yielding a formal
limit of 
$ \tau \ls 0.12 $~hr (although this paper adopts a more conservative limit
of $ \tau \ls 0.3 $~hr), while the Interval~A VS/VH CCF is essentially
unusable because it has two poorly defined, nearly equal peaks of $r_{max}
\approx 0.57 $.}
\label{fig3}
\end{figure}

\clearpage


\begin{references}

\reference{B89} Bregman, J.\ 1990, Astronomy and Astrophysics Review (ISSN
0935-4956), v.\ 2, no.\ 2, p.\ 125-166
\reference{C99} Chiappetti, L.\ \et 1999, ApJ, 521, 552
\reference{E89} Edelson, R., Krolik, J.\ 1989, ApJ, 333, 646
\reference{E95} Edelson, R.\ \et 1995, ApJ, 438, 120
\reference{F00} Fossati, G.\ \et 2000, ApJ, in press (astro-ph/0005066)
\reference{G97} Ghisellini, G.\ \et 1998, MNRAS, 301, 451
\reference{J00} Jansen, F., Lumb, D., Altieri, B., \et 2001, A\&A, 365, in
press
\reference{K00} Kataoka, J.\ \et 2000, ApJ, 528, 243
\reference{M97} Mastichiadis, A., Kirk, J.\ 1997, A\&A 320, 19
\reference{P98} Peterson, B.\ M.\ \et 1998, PASP, 110, 660
\reference{S97} Sikora, M.\ \et 1997, ApJ, 484, 108
\reference{S00} Str\"{u}der, L., Briel, U. G., Dennerl, K., \et 2001,
A\&A,
365, in press
\reference{T96} Takahashi, T.\ et 1996, ApJL, 470, L79
\reference{T00} Takahashi, T.\ et 2000, ApJ, in press (astro-ph/0008505)
\reference{T00} Turner, M. J. L., Abbey, A., Arnaud, M., \et 2001, A\&A,
365, in press (astro-ph/0011498)
\reference{U97} Urry, C.\ M.\ \et 1997, ApJ, 486, 799
\reference{W94} White, R., Peterson, B.\ M.\ 1994, PASP, 106, 879
\reference{U96} Zheng, Y.\ \et  1999 ApJ, 527, 719

\end{references}
\end{document}